\def\beq{\begin{equation}}
\def\eeq{\end{equation}}
\def\bea{\begin{eqnarray}}
\def\eea{\end{eqnarray}}

\documentclass[aps,
               floatfix,
               showpacs,
               showkeys,
               superscriptaddress]{revtex4}  
\usepackage{isolatin1}
\usepackage{graphics,graphicx}
\usepackage{amsmath}
\usepackage{amssymb}

\begin{document}
\title{Escape patterns due to ergodic magnetic limiters in tokamaks
               with reversed magnetic shear}

\author{Marisa Roberto}
\affiliation{Instituto Tecnológico de Aeronáutica,
             Centro Técnico Aeroespacial, Departamento de Física,
             12228-900, São José dos Campos, São Paulo, Brazil.}

\author{Elton C. da Silva}
\affiliation{Instituto de Física, Universidade de São Paulo,
             C.P. 66318, 05315-970, São Paulo, São Paulo,
             Brazil.}

\author{Iberê L. Caldas}
\affiliation{Instituto de Física, Universidade de São Paulo,
             C.P. 66318, 05315-970, São Paulo, São Paulo,
             Brazil.}

\author{Ricardo L. Viana}
\affiliation{Departamento de Física, Universidade Federal
             do Paraná, C.P . 19081, 81531-990, Curitiba, Paraná,
             Brazil.}

\date{\today}   

\begin{abstract}

In this work we study the ergodic magnetic limiters (EML) action on field lines from the point of view of a chaotic scattering process, considering the so-called exit basins, or sets of points in the chaotic region which originate field lines hitting the wall in some specified region. We divide the tokamak wall into three areas of equal poloidal angular length, corresponding to different exits for a chaotic field line. In order to obtain the exit basins we used a grid chosen inside a small rectangle which comprises a representative part of the chaotic region near the wall. Thus, exit basins were obtained for a tokamak wall with reversed magnetic shear. The nontwist mapping describes the perturbed magnetic field lines with two chains of magnetic islands and chaotic field lines in their vicinity. For a perturbing resonant magnetic field with a fixed helicity, the observed escape pattern changes with the perturbation intensity. 
\end{abstract}
\maketitle

\section{Introduction}

Area preserving maps have been used to describe magnetic field line
behaviour in fusion plasma confinement schemes like tokamaks
\cite{chirikov} \cite{wesson}. The
magnetic field line equations, after proper parametrization (the
time-like variable is an ignorable coordinate), can be viewd as
Hamilton's equation describing an integrable systems, where phase
space trajectories are identified with magnetic field lines
\cite{morrison}. If the torus is intersected by a Poincar\'{e} surface
of section we can write down an area preserving canonical mapping
relating the coordinates of a field line at a given intersection in
terms of the same coordinates at a previous time.

The use of two-dimensional sympletic maps enables us to harness the
powerful methods of Hamiltonian dynamics in order to explain the
observed changes in the magnetic field structure due to magnetostatic
perturbations \cite{lichtenberg} \cite{petrisor} \cite{punjabi} \cite{abdu}. In this paper we create an outer
chaotic layer of magnetic field lines superimposing the tokamak
equilibrium field to magnetic field generated by an ergodic magntic
limiter (EML), which consists of slices of external helical conductors
with a suitable pitch \cite{caldas} \cite{karger}. The equilibrium field presents
reversed reversed magnetic shear, which has been proposed as a way to
obtain transport barriers which can help improving the confinement
quality \cite{roberto} \cite{howard}.

In this work we study the EML action on field lines from the point of view
of a chaotic scattering process \cite{elton_basins} \cite{bleher}. The point on the tokamak wall where a  given chaotic field line will eventually hit depends sensitively on which point on the wall it cames from. Hence it is useful to consider the so-called exit basins, or sets  of points in the chaotic region which originate field lines hitting the wall in some specified region. The mathematical structure underlying the chaotic region is an extremely involved tangle comprised of homoclinic intersections between invariant manifolds of unstable orbits embedded in the chaotic region. Due to this structure we expect that the exit basins have a nontrivial geometric structure. 

In Sec. II we present the equilibrium and perturbing magnetic field. The exit basins for the EML map are obtained in Sec. III and in the last section we present our conclusion.

\section{Equilibrium and perturbing magnetic fields}

Many coordinate systems have been used to describe magnetic field geometry in plasma confinement systems. In this paper we will work with the non-orthogonal coordinates $(r_t,\theta_t,\varphi_t)$, which resemble the commonly used toroidal coordinates $(\xi,\omega,\Phi)$  but present the Shafranov shift in the
``right'' (inward) direction \cite{kucinski}. In the large aspect ratio limit
they reduce to the pseudo-toroidal, or local coordinates
$(r,\theta,\Phi)$.

The tokamak equilibrium magnetic field ${\bf B}_0$ is obtained from an approximated analytical solution of the Grad-Schl\~uter-Shafranov equation in these coordinates \cite{mutsuko90}: 
\beq
\label{gradsha}
\Psi_p(r_t,\theta_t) = \Psi_{p0}(r_t) + \delta \Psi_p(r_t,\theta_t), 
\eeq
\noindent where
\beq
\label{psi0}
\frac{d \Psi_{p0}(r_t)}{d r_t} = \frac{\mu_0 I_p R_0^\prime}{2\pi r_t} 
\left\lbrack 1 - \left( 1 + \beta' \frac{r_t^2}{a^2} \right)
{\left( 1 - \frac{r_t^2}{a^2} \right)}^{\gamma + 1} \right\rbrack,
\eeq
\noindent with $a$ as the plasma radius, determined by a material limiter, $\beta' \equiv \beta (\gamma+1)/(\beta+\gamma+2)$, where $\beta$ and $\gamma$ are positive parameters, and $|\delta \Psi_p(r_t,\theta_t)| \ll |\Psi_{p0}(r_t)|$. $I_p$ is the total plasma current and $R_0^\prime$ is the major radius of the circular centre (i.e., the magnetic axis radius).

In the large aspect ratio limit, and supposing that in lowest order the equilibrium flux function $\Psi_p(r_t)$ does not depend on $\theta_t$, the Grad-Schl\~uter-Shafranov equation reduces to an equilibrium equation similar to that obtained in a cylindrical system, but now in terms of $r_t$. The intersections of the flux surfaces $\Psi_p(r_t) = constant$ with a toroidal plane are not concentric circles but rather present a Shafranov shift toward the exterior equatorial region. Hence, actual equilibrium flux surfaces can be approximated by $r_t = const.$ coordinate surfaces. In the common range of tokamak parameters, as those considered in this paper, the aspect ratio is always large enough to ensure that the $r_t = const.$ surfaces do not overlap. 

We have used a toroidal current density profile with a central hole, given by \
\beq
\label{j30}
J_{3}(r_t) = \frac{I_p R_0^\prime}{\pi a^2}
\frac{(\gamma+2)(\gamma+1)}{\beta+\gamma+2}
\left( 1 + \beta \frac{r_t^2}{a^2} \right)
{\left( 1 - \frac{r_t^2}{a^2} \right)}^\gamma.
\eeq
\noindent The safety factor profile has a non-monotonic profile, what accounts for describing the reversed shear effect. For some values of the safety factor there are two magnetic surfaces with different radii within the plasma column. 

\begin{figure}
\includegraphics[width=0.4\textwidth]{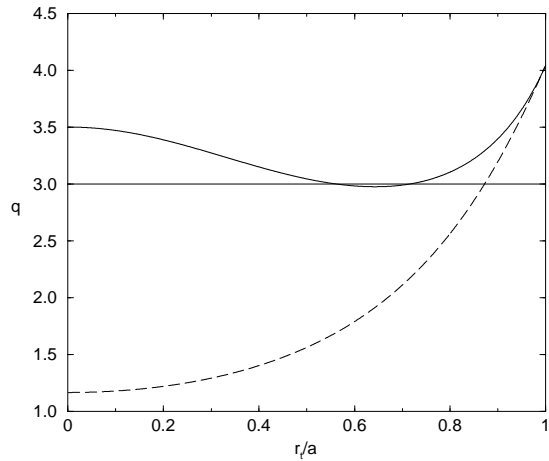}
\caption{\label{equilibrio} Safety factor profile for equilibria with $\gamma = 1.0$, $\beta = 3.0$ (solid line), and $\gamma = 2.0$, $\beta = 0.0$ (dashed line).}
\end{figure}

In the numerical simulations to be described in this paper, we normalize the minor tokamak radius $b_t$, and the plasma radius $a$ to the major (magnetic axis) radius ${R'}_0$, such that $a/{R'}_0 = 0.25$ and $b/{R'}_0 = 0.33$. We also choose $q(a) = 4.04$ and $q(0) = 3.50$, corresponding to the safety factors at the plasma edge and magnetic axis, respectively, as observed in typical discharges with negative magnetic shear, for which $\beta = 3.0$ and $\gamma = 1.0$. Figure \ref{equilibrio} depicts the corresponding radial profile of the safety factor (solid line). For comparison, a usual monotonic radial profile for $q(0) = 1.25$ and the same value of $q(a) = 4.04$ is also shown in Fig. \ref{equilibrio}. 

\begin{figure}
\includegraphics[width=0.5\textwidth]{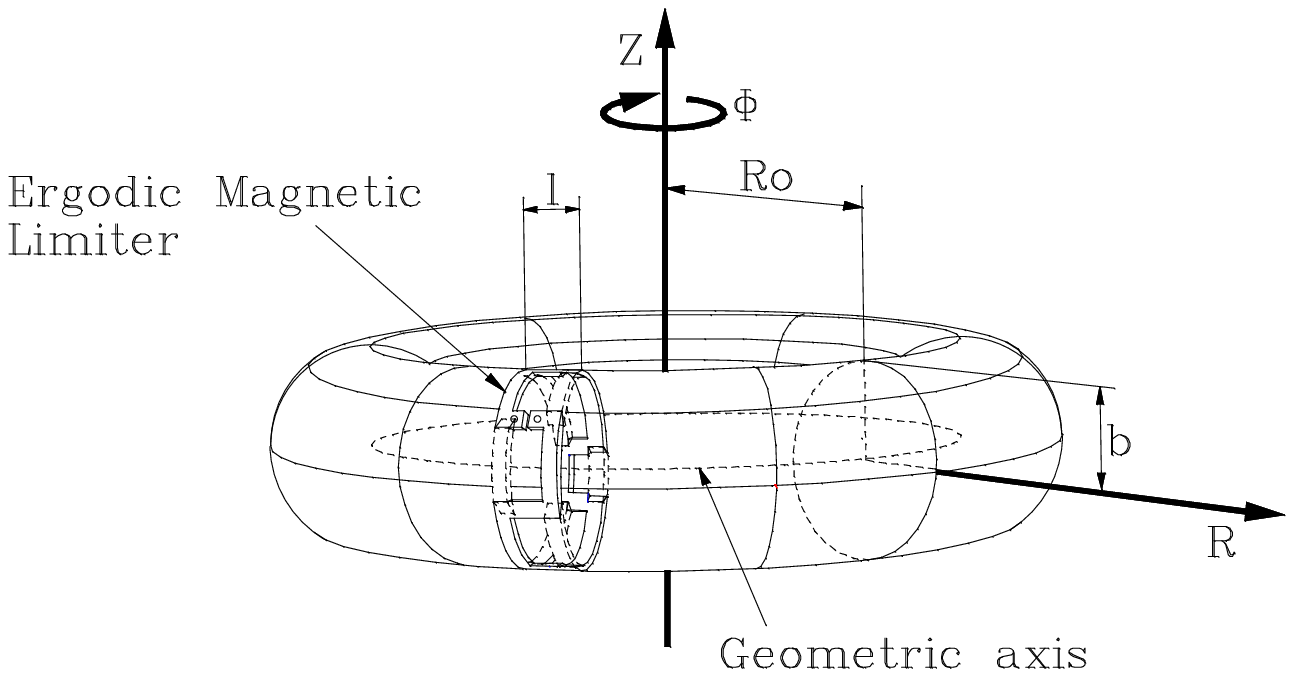}
\caption{\label{limitador}Scheme  of an ergodic magnetic limiter.}
\end{figure}

The design for the ergodic magnetic limiter to be considered in this paper is essentially the same as in Ref. \cite{caldas}, and consists of $N_r$ current rings of length $\ell$ located symmetrically along the toroidal direction of the tokamak (Figure \ref{limitador}). These current rings may be regarded as slices of a pair of external helical windings located at $r_t = b_t$, conducting a current $I_h$ in opposite senses for adjacent conductors. The role of these windings is to induce a resonant perturbation in the tokamak, and to achieve this effect we must choose a helical winding with the same pitch as the field lines in the rational surface we want to perturb. This has been carried out by choosing the following winding law $u_t = m_0 \theta_t - n_0 \varphi_t =$ constant. In this paper we will consider an ergodic limiter consisting of $N_r = 4$ rings with mode numbers $(m_0,n_0)=(3,1)$ each, carrying a current $I_h$.

\begin{figure}
\includegraphics[width=0.4\textwidth]{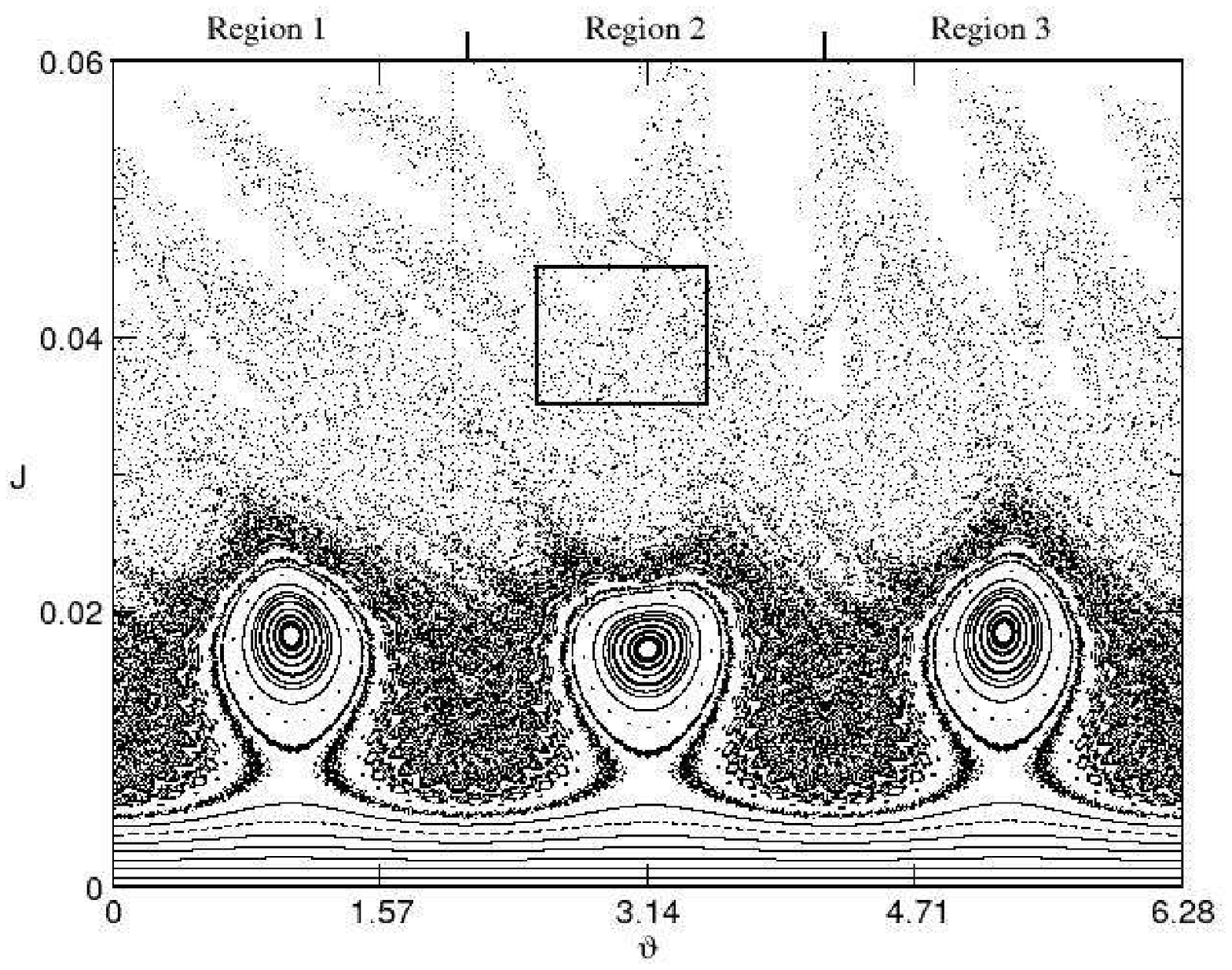}
\includegraphics[width=0.4\textwidth]{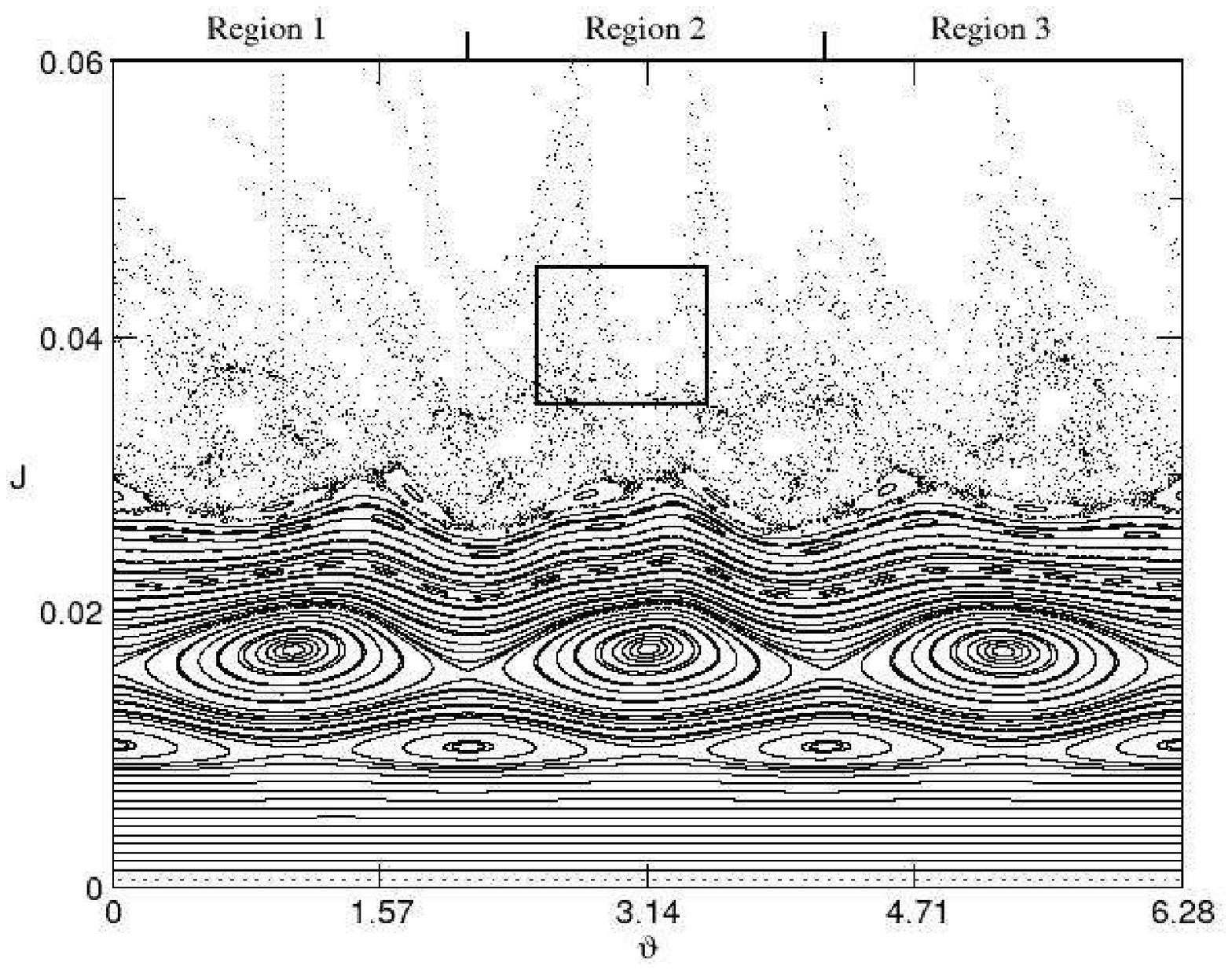}
\caption{\label{poincare} Poincar\'e maps for the ergodic limiter mapping, an equilibrium with  $\beta = 3.0$ and $\gamma = 1.0$, and normalized limiter currents $I_h/I_p = $ (a) $7.0 \%$ with $m_0 = 3$  and  (b) $7.0 \%$ with $m_0 = 5$.}
\end{figure}

We can derive, due to the perturbation of the EML, a stroboscopic map for field line dynamics in terms of ${\cal J}_n$ and $\vartheta_n$ as the action and angle variables. In Figures \ref{poincare}(a) and \ref{poincare}(b) we  show phase portraits for a limiter current for $I_h/I_p = 7.0 \%$. We have chosen a limiter with $m_0 = 3$ and $m_0 = 5$  pairs of current wires, respectively, its perturbing field resonates with the equilibrium tokamak field and generates chains of three magnetic islands. Since the safety factor radial profile is non-monotonic, for $q = 3.0$ there are two distinct radial locations at which there are such chains [see Fig. \ref{equilibrio}].

\section{Exit basins for the EML map}

Suppose we choose at random a magnetic field line with initial
condition located outside the last closed magnetic surface, i.e. in
the chaotic region near the tokamak wall. We can think of it as a
trajectory that eventually goes to the tokamak wall by iterating the
mapping forward to $n \rightarrow \infty$. This occurs because the
wall at $r_t = b_t$ is actually an arbitrary partition in the phase
portraits, and the chaotic region intersects the line $r_t = b_t$ at a
finite line segment. In terms of the analogy with chaotic scattering process, we may consider these field lines as {\it outgoing} trajectories. Since the EML map is invertible, the same initial conditions generate field lines that eventually hit the wall as $n \rightarrow -\infty$ when we iterate the map backwards. Accordingly, we consider them as {\it ingoing} trajectories. 

While for dissipative systems we speak of basins of attraction to refer to the set of initial conditions which converge to a given attractor, it is possible to extend this concept for including the set of initial conditions which generates trajectories that escape through a given exit, in an open Hamiltonian system \cite{bleher}. This set is called an {\it exit basin}. When there are two (or more) exits in the system one is interested in the exit basin boundary, which can be either smooth or fractal, as for basins of attraction of dissipative systems. Fractal boundaries are important dynamical objects because orbits that start in their vicinity exhibit very complicated and unpredictable motion. In our case, we divide the tokamak wall into three poloidal sections of equal length, $0 \le \vartheta < 2\pi/3$, $2\pi/3 \le \vartheta < 4\pi/3$, and $4\pi/3 \le \theta < 2\pi$, indicated respectively as regions 1, 2, and 3 corresponding to different exits for a chaotic field line. 

In order to obtain the exit basins we used a fine grid of $600\times
700$ points chosen inside the small rectangle shown in Figure
\ref{poincare}(a) and \ref{poincare}(b), and which comprises a
representative part of the chaotic region near the wall. We mark the
initial condition pixel in dark gray, light gray or black, depending
on whether the field line goes to the wall section called region~1, ~2
or ~3, respectively, under the forward dynamic. Figures
4a and 4b show the exit basins for the
maps showed in Figures 3a and 3b for non-monotonic safety factor
profile. According  to the Figure
3a the escape concentrates on the external equatorial region if the
internal modes (3,1) are perturbated. If we consider external modes as
(5,1), the field lines spread over all studied regions. Figure
\ref{WBac} shows the exit basins for the monotonic safety factor
profile. It can be seem that the exit basins
are intertwined in a very complex way for all cases. We conjecture that
the exit basin boundary in Figures 4b and 5 are fractals. 

\begin{figure}
\includegraphics[width=0.4\textwidth]{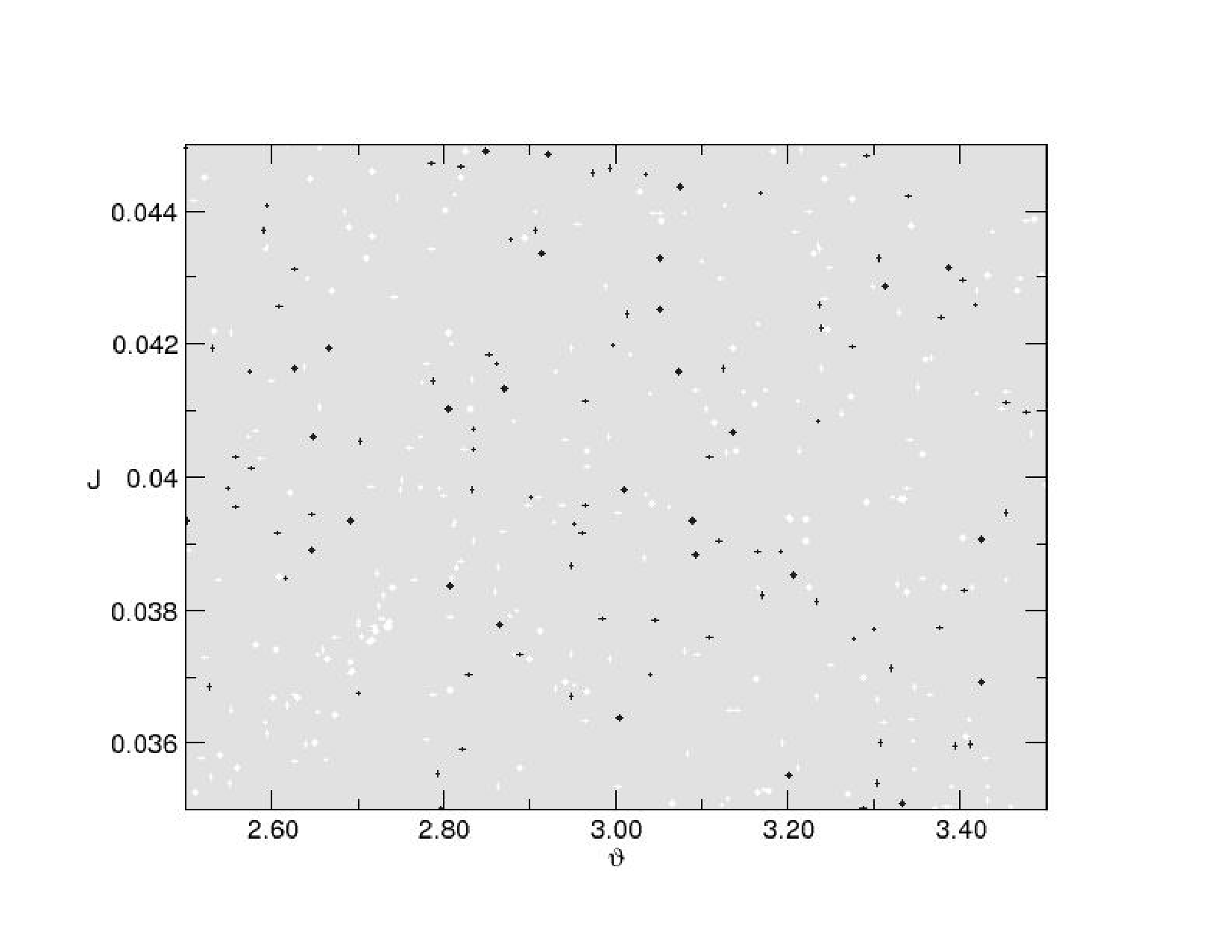}
\includegraphics[width=0.4\textwidth]{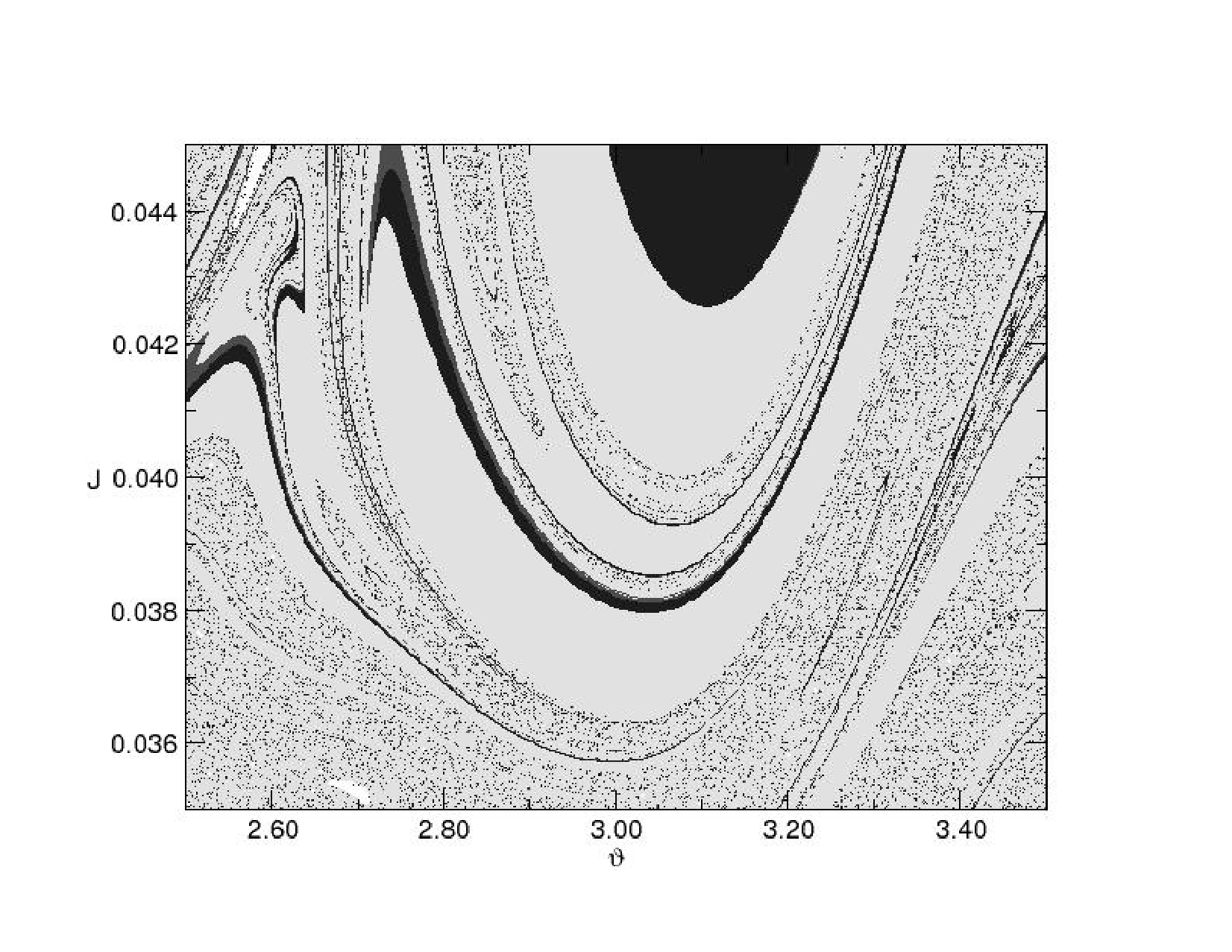}
\caption{\label{Bacias} Exit basin for the EML maps showed in Figures
3a and 3b. The regions in
dark gray, light gray, and black correspond to field lines colliding
with the tokamak wall at the regions marked 1, 2, and 3, respectively
in Figure \ref{poincare}. The field lines that do not escape are
marked in white color.
}
\end{figure}

\begin{figure}
\includegraphics[width=0.4\textwidth]{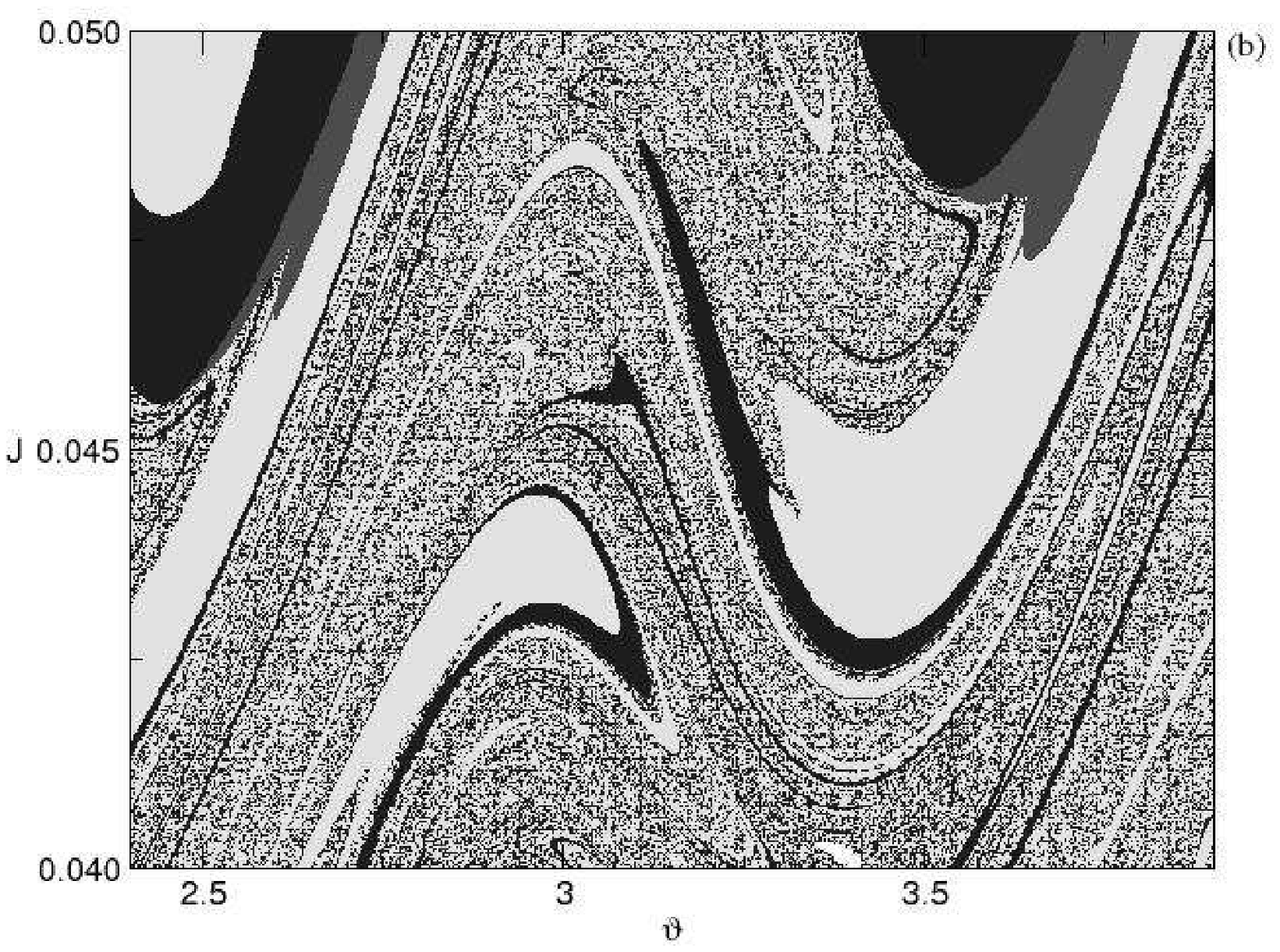}
\caption{\label{WBac}} Exit basin for the EML map, for monotonic radial safety factor profile, shown in Fig. \ref{equilibrio}. 
\end{figure}

\section{Conclusions}

 The creation of a chaotic field line region near the tokamak wall 
has been proposed as a way to uniformize heat and particle loadings on
 the wall, so diminishing plasma contamination. However, the available
 experimental results point out that this may be not necessarily the 
case, since particle deposition is not uniform, as originally thought. 
Our work aimed to shed some light on this matter by considering the 
dynamical structure of the escape channels {\it inside} the chaotic 
region. The existence of escape channels for magnetic field lines
 make the particle deposition non-uniform, as long as we neglect 
particle drifts. On the other hand, the escape channels are closely 
related to a dynamical property of the open chaotic system named 
exit basins, which have been described qualitatively in this work. 
Since the exit basin boundary structure is very complex, presenting 
a fractal nature, we conclude that the field line impacts on the
 tokamak wall are correspondingly complex, presenting also a fractal
 characteristic.

\section*{Acknowledgments}

This work was made possible with partial financial support of the following
agencies: CNPq (Conselho Nacional de Desenvolvimento Cient\'{\i}fico e
Tecnol\'ogico), CAPES (Coordena\c c\~ao de Aperfei\c coamento de Pessoal de
N\'{\i}vel Superior), FAPESP (Funda\c c\~ao de Amparo \`a Pesquisa do Estado
de S\~ao Paulo) , and Funda\c c\~ao Arauc\'aria (State of Paran\'a, Brazil).  



\begin{thebibliography}{99}

\bibitem{chirikov} B. Chirikov, Phys. Reports {\bf 52}, 265 (1979). 
\bibitem{wesson} J. Wesson, Tokamaks (Cambridge University Press, 1987).
\bibitem{morrison} P.J. Morrison, Phys. Plasmas {\bf 7}, 2279 (2000).
\bibitem{lichtenberg} A.J. Lichtenberg and M.A. Lieberman, Regular and
Chaotic Dynamics, 2nd. Edition (Springer-Verlag, New
York-Berlin-Heidelber, 2002).
\bibitem{petrisor} E. Petrisor, J. H. Misguisch, D. Constantinescu, Chaos,
Solitons and Fractals, {\bf18}, 1085 (2003).
\bibitem{punjabi} Ali H, Punjabi A. Boozer A, Evans T, Phys. Plasmas {\bf11}, 1908 (2004).
\bibitem{abdu} S.S.Abdullaev, Nuclear Fusion {\bf44}, S12 (2004).
\bibitem{caldas} E.C. Silva, I.L. Caldas, R.L.Viana, Phys. of Plasmas,
{\bf8}, 2855 (2001).
\bibitem{karger} F. Karger and F. Lackner, Physics Letters A 61
  385 (1977). 
\bibitem{roberto} M. Roberto, E.C. Silva, I.L. Caldas and R.L. Viana,
Phys. Plasmas {\bf 11}, 214 (2004).
\bibitem{howard} J. E. Howard and S. M. Hohs, Phys. Rev. A {\bf 29},
418 (1984).
\bibitem{elton_basins} E. C. Silva, I. L. Caldas, R. L. Viana and Miguel A.F. Sanjuan, Phys. Plasmas {\bf 9}, 4917 (2002).
\bibitem{bleher} S. Bleher, C. Grebogi, E. Ott, and R. Brown, Phys. Rev. A {\bf 38}, 930 (1988).
\bibitem{kucinski} M. Y. Kucinski and I. L. Caldas, Z. Naturforsch. {\bf 42a},
  1124 (1987). 
\bibitem{mutsuko90} M. Y. Kucinski, I. L. Caldas, L. H. A. Monteiro, and
  V. Okano, J. Plas. Phys. {\bf 44}, 303 (1990). 

\end{thebibliography}
\end{document}